# Mechanical Creep Instability of Nanocrystalline Methane Hydrates

Pinqiang Cao [1, *], Jianlong Sheng [1], Jianyang Wu [2], Fulong Ning [3],

[1] School of Resource and Environmental Engineering, Wuhan University of Science and Technology, Wuhan, Hubei 430081, China

[2] Department of Physics, Jiujiang Research Institute, Research Institute for Biomimetics and Soft Matter, Fujian Provincial Key Laboratory for Soft Functional Materials Research, Xiamen University, Xiamen 361005, China

[3] Faculty of Engineering, China University of Geosciences, Wuhan, Hubei 430074, China

Email address:

Pinqiang Cao: pinqiang@wust.edu.cn;

Jianlong Sheng: shengjl@wust.edu.cn;

Jianyang Wu: jianyang@xmu.edu.cn;

Fulong Ning: nflzx@cug.edu.cn;

*To whom correspondence should be addressed: pinqiang@wust.edu.cn;

**Abstract:** Mechanical creep behaviors of natural gas hydrates (NGHs) are of importance for understanding mechanical instability of gas hydrate-bearing sediments on Earth. Limited by the experimental challenges, intrinsic creep mechanisms of nanocrystalline methane hydrates remain largely unknown yet at molecular scale. Herein, using large-scale molecular dynamics (MD) simulations, mechanical creep behaviors of nanocrystalline methane hydrates are investigated. It is revealed that mechanical creep responses are greatly dictated by internal microstructures of crystalline grain size and external conditions of temperature and static stress. Interestingly, a long steady-state creep is observed in nanocrystalline methane hydrates, which can be described by a modified constitutive Bird-Dorn-Mukherjee model. Microstructural analysis show that deformations of crystalline grains, grain boundary (GB) diffusion and GB sliding collectively govern the mechanical creep behaviors of nanocrystalline methane hydrates. Furthermore, structural transformation also appears important in their mechanical creep mechanisms. This study sheds new insights into understanding the mechanical creep scenarios of gas hydrates.

**Introduction**

Natural gas hydrates (NGHs) are considered as an important potential energy resource because of vast amounts of carbon resource (*e.g.*, methane, ethane) stored in gas hydrate reservoirs [1-3]. NGHs are non-stoichiometric solid crystalline compounds with a host lattice of water molecules enclathrating suitable guest gas molecules at high pressures and low temperatures [4, 5]. To date, three common crystalline structures of gas hydrates have been identified, including structural I, II, and H hydrates. In molecular structural point of view, those gas hydrates are characterized by distinct arrangements of several polyhedral water cavities formed by tetrahedral hydrogen (H)-bonds. The structural stability of these polyhedral water cavities can be promoted by accommodating a variety of suitable guest molecules, *e.g.*, $CH_4$, $C_2H_6$, $CO_2$, $H_2$, $H_2S$, Xe, Kr, and so on. Although gas hydrates are not as much more commonly experienced as ice on a daily basis, gas hydrates have drawn a lot of attention in the world due to their significant role in flow assurance [4, 6], global climate change [7-9], and gas storage or transportation [10-12].

Gas hydrates are extensively distributed in submarine continental margins and permafrost regions [13-15], even on outer planets in the Universe [16-18]. Of particular concern is one fact, which gas hydrates are very sensitive to changes of external conditions such as temperature and pressure, resulting from the natural environmental changes and human activities. In natural environment, complex stratums in Earth interior are throughout dynamical, accompanied by geological movement over million years. As a result, gas hydrate-bearing sediments in natural settings may be always undergoing mechanical creep deformation under geological processes. Understanding creep deformation of nanocrystalline gas hydrates can give strong implications for assessing the mechanical instability of gas hydrate-bearing sediments on Earth. Such knowledge about gas hydrates is the basis of gas hydrate exploitation and utilization. To data, some experimental studies have been performed to investigate the mechanical creep behaviors of gas hydrates under static stress. For example, Durham *et al*.[19, 20] reported via compressional creep tests that nanocrystalline methane hydrates are extraordinarily strong compared to other icy compounds. The different mechanical behavior of both nanocrystalline methane hydrates and ice can be attribute to the different coordinated motion of crystalline defects typically rate-limited by diffusion and the size difference in unit cell of hydrate and ice crystals [19-21]. Using triaxial compression tests conducted on methane-hydrate-bearing sand, Miyazaki *et al*.[22] found that methane-hydrate-bearing sediment specimens exhibit strong time-dependence. Moreover, it was revealed that, based on a simple hypothesis in the constant-stress/strain-rate tests, the creep life correlates with the loading-rate dependencies of strength[22].

Despite important progress in experimental studies of creep responses of gas hydrates, limited by experimental operation challenges and meta-stability of nanocrystalline samples at conventional external conditions, the fundamental knowledge of creep behaviors of methane hydrates still remains largely insufficient at the nanoscale, particularly for intrinsic creep mechanisms. This limits the evaluation of stability of NGHs reservoirs under external activities from a long-period insight. Therefore, large-scale molecular dynamics (MD) simulations are performed to reveal the nature of physical creep phenomenon and elucidate microscopic creep characteristics of nanocrystalline methane hydrates from an atomistic perspective. In this work, the effects of grain size, external static stress and temperature on the creep behaviors of nanocrystalline methane hydrates are investigated, respectively.

**Methods**

**Molecular Models of Polycrystalline Methane Hydrates**

The initial configurations of polycrystalline methane hydrate samples are constructed by Voronoi tessellation method [23]. As-constructed samples are textured microstructures that consist of a number of nanograins of structural I methane hydrate characterized by $5^{12}$ and $5^{12}6^2$ clathrate cages, as illustrated in Figure 1. To examine the effects of grain size on the creep behaviors, five polycrystals with an average grain size ranging from approximately 4 to 16 nm are constructed, and their dimensions vary from about $9.3 \times 9.3 \times 9.3$ to $37.2 \times 37.2 \times 37.2$ nm$^3$. Each polycrystal contains 24 Voronoi nanograins with randomly crystalline orientations and polygonal shapes. This agrees with the fact that, via powder *X-ray* analysis, laboratory-prepared methane hydrate samples do not show preferred crystallographic orientations of polycrystalline grains [24]. As a result, each polycrystal show unique microstructures of grain boundaries (GBs) and grain junctions (GJs). To avoid artificial close contact and overlap of molecules in each polycrystal, molecules protruding beyond GBs and GJs are removed when part of a molecule pair with a nearest neighbour distance is below 0.1 nm. The as-constructed samples contain approximately 27,729 to 1,771,683 molecules, depending on the dimension. To achieve statistical analysis for polycrystalline methane hydrate with a given grain size, six samples with different grain orientations and grain shapes are constructed. Periodic boundary conditions (PBCs) are imposed in the three orthogonal directions to mimic infinite polycrystalline structures for thermodynamic limits.

**Forcefield for Modeling Polycrystals**

Both host water and guest methane molecules are described using monatomic model that takes each molecule as a coarse-grained spherical particle. To describe the intermolecular interactions in polycrystalline methane hydrates, the Stillinger-Weber potential [25] parameterized by Molinero *et. al*, [26, 27] is utilized. The Stillinger-Weber forcefield [25] is able to mimic the tetrahedral short-ranged interactions of monatomic water (mW), as well as intermolecular water-methane and methane-methane interactions [26, 27]. This coarse-grained forcefield is approximately over 2-orders of magnitude more efficient than fully atomistic models in reproducing a range of properties of liquid- and solid-stated water. Moreover, it also has been successfully utilized to predict the mechanical characteristics of methane hydrates [28].

**MD Creep Simulations**

Initially, as-created polycrystalline samples are quasi-statically relaxed to a local minimum energy configuration using the conjugate gradient method with an energy tolerance of $1.0 \times 10^{-4}$ eV and a force tolerance of $1.0 \times 10^{-4}$ eV/Å, respectively. Then, MD relaxation with a simulation time of 1 ns is performed at given temperature and pressure of 10 MPa under NPT (constant number of particles, constant pressure, and constant temperature) ensemble. The confining pressure and temperature are controlled using Nosé-Hoover barostat and thermostat with damping time constant $\tau_T = 0.1$ ps and $\tau_p = 1$ ps, respectively. Thereafter, another MD relaxation with a simulation time of 1 ns is performed at given temperature under NVT (constant number of particles, constant temperature, and constant volume) ensemble using Nosé-Hoover thermostat with damping time constant $\tau_T = 0.1$ ps. Finally, creep MD simulations are performed. To achieve creep tests, a given constant stress is imposed on one orthogonal (*z*) direction, while the confining pressures in the other lateral two (*x* and *y*) directions are controlled at 10 MPa. A timestep of 1 fs with the velocity-Verlet algorithm is used to integrate the Newton equation of atomic motion. The particle stress is calculated according to the virial definition of stress by means of using the forces on the particles collected during the MD calculations. To eliminate the fast thermal fluctuations, both potential energy and stress of particles in polycrystals are averaged over 5000 timesteps. In our work, the half-cage order parameter (H-COP) with a cutoff distance of 3.2 Å is used to detect water cages in polycrystalline methane hydrates[29, 30]. All large-scale MD simulations of polycrystalline methane hydrates are performed using Large-scale Atomic/Molecular Massively Parallel Simulator (LAMMPS) package[31].

**Results**

**Grain Boundary Structures in Polycrystals**

Figure 2 presents side-viewed snapshots of polycrystalline methane hydrates with two different grain sizes at the initial and relaxed states, where the color code is on the basis of potential energy. Similar to previous studies[28, 32], molecules located at grain boundaries (GBs) and grain boundary junctions (GJs) in polycrystals show higher potential energies than those at grain interiors (GIs), as shown in Figures 2a-2f. This indicates that GBs and GJs are metastable structures. MD relaxations significantly reduce the potential energies of GBs and GJs in polycrystals; however, the potential energies of GBs and GJs are still higher than those at GIs. Interestingly, analysis of molecular structures shows that there are several unusual types of water polyhedral water cages identified at the GBs and GJs, which is similar to the previous studies[28].

**Mechanical Creep Behavior of Polycrystals**

**Creep Behavior in Polycrystals**

Figure 3 shows the creep strain-time curves of polycrystalline methane hydrates with different mean grain sizes at different temperatures and subjected to different static stresses. As shown in Figure 3a, three mechanical creep stages can be roughly identified from the creep responses of polycrystalline methane hydrates, indicating that deformation creep mechanisms may change with increasing creep time. The first primary creep deformation stage is primarily characterized by nonlinear responses in the creep strain, where creep strain becomes less pronounced with increasing creep time. This primary creep stage is very short duration. The secondary creep stage is characterized by linearity in the creep strain responses, indicating a steady-state of creep under constant load. By comparison, the creep time of the secondary stage is much longer than that of the primary creep stage. The last tertiary creep stage is described by significant increase of creep strain with the increasing creep time, suggesting rapid creep deformation and final mechanical fracture of polycrystalline methane hydrates.

**Grain Size-Dependence**

As is known, crystalline grain size plays an important role in the mechanical strength of polycrystalline methane hydrates and ice[28, 33] and other conventional polycrystalline solid materials[34-36]. To date, however, the effects of grain size on mechanical creep behaviors of polycrystalline methane hydrates remain unexplored. Here, the critical role of grain size on the creep behaviors of polycrystalline methane hydrates is examined for the first time (Figure 3b and Figure S1). Figure 3b shows MD simulated average creep strain-time curves of polycrystalline methane hydrates with grain sizes varying from around 4.0 to 16.0 nm under a constant static stress. Clearly, polycrystalline methane hydrates with different grain sizes show distinct creep strain-time curves, indicating strong grain size dependence on creep deformation. Intriguingly, there is critical grain size (around 10 nm) in the creep behaviors of polycrystalline methane hydrates. Below which, it is identified that the creep strain becomes more pronounced with the increase of stress-loading time at a given static stress. This indicates that, under a constant load, the overall creep deformation increases with reduction in the grain size of polycrystalline methane hydrates. This is explained by the fact that polycrystals with small grains show high density of weak GBs and GJs that play a critical role in the mechanical creep behaviors of nanocrystalline methane hydrates. Unlike bulk methane hydrates, GBs and GJs are semi-crystalline weak structures, in which water and methane molecules fast diffuse, resulting in their distinct creep behaviors. This is resembling those of other engineering materials[37]. Above which, however, the creep strain is also higher than that of polycrystalline methane hydrates with critical grain size of about 10 nm at a given time. This interesting behavior of polycrystalline methane hydrates can be attributed to the large-scale collective GB sliding (Figure 4g-4i). Such critical grain size has been also detected in the mechanical strength of polycrystalline methane hydrates from previous studies[28]. This implies that the mechanical stability of gas hydrate-bearing sediments is strongly dictated by the grain size of polycrystalline gas hydrates.

**Stress-Dependence**

According to the classical constitutive Bird-Dorn-Mukherjee equation[38], the external static stress plays a key role in the mechanical creep process of materials. Figures 3c-e show the typical creep strain-time curves of nanocrystalline methane hydrates with mean grain size of 4.0, 7.0 and 10.0 nm subjected to uniaxial loading stress varying from 100 to 300 MPa, respectively. Apparently, the mechanical creep responses strongly correlate with the static loading stress. For a given mean grain size, it is observed that nanocrystalline methane hydrates show higher creep strain subjected to larger static loading stress at a given creep time, indicating that larger creep strain occurs as a higher static stress is applied. Furthermore, the creep strain is nonlinearly increased with the increase of static loading stress. This indicates that gas hydrate-bearing sediments are more easily destabilized as large external forces are imposed. Moreover, this also suggests that they show different mechanical stabilities when they occur in different ground depths. Such stress-dependence of creep behaviors originates from the fact that molecules more easily overcome the energy barrier at higher static loading stress level, resulting in faster diffusion rate of molecules, which is in agreement with previous studies[21].

**Temperature-Dependence**

The effects of temperature on the mechanical creep behaviors of nanocrystalline methane hydrates are also investigated. Figures 3f-h present the typical creep strain-time curves of nanocrystalline methane hydrates with average grain size of 4.0, 7.0 and 10.0 nm at temperatures varying from 203.15 to 263.15 K under constant external static loading stresses, respectively. It is identified from Figure 3f that there is a negligible effect on the creep responses of polycrystalline methane hydrates with mean grain size of about 4.0 nm as the temperature is below 223.15 K, while a strong positive effect is detected when the temperature is above 223.15 K. However, polycrystalline methane hydrates with large mean grain size show a strong positive effect of temperature on the creep strain under constant stress of 300 MPa. As the temperature is increased from 203.15 to 263.15 K, the mechanical creep strain is nonlinearly enlarged. This creep phenomenon can be attributed to the high vibration frequency of water and methane molecules at high temperature, resulting in that molecules are more likely to diffuse along GBs and GJs at high temperature. This suggests that gas hydrate-bearing sediments are more easily destabilized as external heat is injected as a result of global warming, petro-activities and so on.

**Microscopic Mechanisms underlying Creep Behaviors**

To in-depth elucidate the mechanical creep mechanisms behind the creep phenomena of polycrystalline methane hydrates, microstructural development of the polycrystals during creep process is recorded. Figure 4 presents the typical snapshots of the localized molecular structures of polycrystalline methane hydrates at different static stress-loading times. Overall, the whole mechanical creep processes of polycrystalline methane hydrates are dominated by a combination of crystalline grain deformation and dissociation, GB diffusion and GB sliding. As indicated by the molecular displacements in Figure 4a-4f, both water and methane molecules diffuse disorderly at a given temperature and static loading stress. Due to the large diffusion distance of water and methane molecules located at the GBs and GJs, GBs are the dominant source of creep GB diffusion in polycrystalline methane hydrates. With the development of GB diffusion, the geometrical shape and morphology of crystalline grains changes to facilitate the creep deformation of polycrystalline methane hydrates as a result of dissociation and reformation of methane hydrates. At large creep deformation strains, GB sliding occurs and facilitates the large deformation (Figure 4g-4i), leading to the rapid development of microstructures in polycrystalline methane hydrates. Those indicates that GB diffusion and GB sliding are critical in dominating the mechanical creep mechanisms of polycrystalline methane hydrates. Interestingly, however, GB sliding becomes more insignificant with decreasing grain size, *e.g.*, 4 nm, although external conditions of temperature and static stress are changed. With regard to polycrystalline methane hydrates with large mean grain sizes, nano-bubbles form at GJs (Figure 4m-4o). Such formation of nano-bubbles enhances the GB diffusion and GB sliding. By comparing localized microstructures at different temperatures (Figures 4a-c and 4j-l), GB diffusion becomes more remarkable at high temperatures. This indicates that the disordered movement of water and methane molecules is enhanced at high temperature. At low temperature (Figure 4j-4l), crystal lattice vibration is also found, and it is comparable to GB diffusion in creep deformation. This suggests that the dominant creep mechanisms of polycrystalline methane hydrates shift from crystal lattice vibration and GB diffusion to GB diffusion/sliding. Subjected to low static stress, the creep mechanisms of polycrystalline methane hydrates are also mainly dominant by crystal lattice vibration and GB diffusion for a given temperature, *e.g.*, 263.15 K (Figure 4m-4o). Interestingly, lattice diffusion is also identified in the creep deformation (Figure S2). One water molecule forming one dodecahedral water cage in hydrate crystals is replaced by another water molecule, and the one water molecule diffuse into other position of hydrate crystals, which can be indicated by the oxygen atom which colored with yellow. This lattice diffusion behavior can be one main source to facilitate the complicated structural transformation of nanocrystalline methane hydrates. Followed by this point above, to further understand the mechanical creep mechanisms of nanocrystalline methane hydrates, quantitative analysis of water cages is preformed to characterize structural transformation of nanocrystalline methane hydrates. As shown in Figure 5, the number of structure I polyhedral $5^{12}$ and $5^{12}6^2$ cages is identified in all nanocrystalline methane hydrates. The high percentages of $5^{12}$ and $5^{12}6^2$ cages suggest that the significant role of the both water cages in the mechanical creep stability of nanocrystalline methane hydrates. Remarkably, the cage percentage of $5^{12}6^2$ cages show dependence on the grain size of nanocrystalline methane hydrates during the entire mechanical creep loads. The higher percentage of $5^{12}6^2$ cages is detected as nanocrystalline methane hydrates have the larger grain size. For polycrystals with small grain sizes (Figures 5a and 5c), the percentage of $5^{12}6^2$ cages is decreasing with the increasing creep strain. In contrast, the percentage of $5^{12}$ cage is increasing with the increasing creep strain. The percentage of $5^{12}$ cages is comparable to the percentage of $5^{12}6^2$ cage at a creep time of around 1.75 ns as the grain size of polycrystals is about 4 nm. This may indicate the unique deformation creep mechanisms of polycrystals with small grain sizes. However, for polycrystals with large grain sizes as shown in Figure 5e, the percentages of $5^{12}$ and $5^{12}6^2$ cages shows nearly constant values with the increasing creep time. Interestingly, several types of uncommon polyhedral water cages, *e.g.*, $5^{12}6^3$, $5^{12}6^4$, $4^15^{10}6^2$, $4^15^{10}6^3$, $4^15^{10}6^4$, and $4^35^66^3$, are also observed, and their changes are much more complicated. For example, the percentage of $5^{12}6^3$ cages on the whole shows an increasing trend with the creep time, especially for polycrystals with large grain sizes. However, the fluctuations of $5^{12}6^3$ cage percentage is much more violent than those of both $5^{12}$ and $5^{12}6^2$ cage percentages. The existence of $5^{12}6^3$ cages suggests that the coexistence of structure I and II gas hydrates based on the previous studies[39, 40]. Besides both $5^{12}$ and $5^{12}6^2$ cage, $4^15^{10}6^2$ cages exhibit the highest percentage in the other identified water cage types during the creep process. Interestingly, the $4^15^{10}6^2$ cages can intermediate the structural transformation between structure I and H gas hydrates[41]. These uncommon polyhedral water cages, *e.g.*, $5^{12}6^3$, $4^15^{10}6^2$, $4^15^{10}6^3$, and $4^15^{10}6^4$, are also observed in the gas hydrate nucleation[40-42]. Therefore, quantitative analysis of water cages indicates structural transformation of nanocrystalline methane hydrates, and structural transformation plays an important role in the mechanical creep mechanisms of nanocrystalline methane hydrates.

**Discussion**

Polycrystalline methane hydrates show apparent three creep deformation stages, namely, primary creep, secondary creep, and tertiary creep. Clearly, the primary creep deformation stage lasts very short time. However, the secondary creep deformation stage is characterized by a steady-state creep for a long time. As for conventional metallic materials, the classical constitutive Bird-Dorn-Mukherjee model[38] was employed to describe steady-state creep as follows:

$$\dot{\varepsilon} = \frac{AD_0Gb}{k_BT}(\frac{b}{d})^p(\frac{\sigma}{G})^n \exp(-\frac{\Delta Q}{k_BT}) \qquad (1)$$

where $\dot{\varepsilon}$ and $k_B$ is the steady-state creep strain rate and the Boltzmann's constant, respectively. $A$, $D_0$, $G$ and $b$ is a dimensionless constant, the diffusion coefficient, the shear modulus and the Burgers vector, respectively. $d$, $T$ and $\sigma$ is grain size, the temperature, and the applied static stress, respectively. $\Delta Q$ is the activation energy in thermal-activated process of substances. As is known, dislocations are mainly identified to dominate the plastic deformation in metallic systems. Whereas, methane hydrates are found to be dislocations-free under large deformation because of the large size of unit cell of hydrates. Here, a modified hydrate-based Bird-Dorn-Mukherjee model is thus developed and proposed to describe steady-state creep as follows:

$$\dot{\varepsilon} = \frac{AD_0 G}{k_B T} (\frac{1}{d})^p (\frac{\sigma}{G})^n \exp(-\frac{\Delta Q}{k_B T}) \qquad (2)$$

Figures 6a and b show logarithmic plots of steady-state creep strain rate versus inverse of grain size under stress-level of 300 MPa and versus stress at 263.15 K, respectively. Upon metallic materials[43-46], both stress and grain size exponents $n$ and $p$ can indicate the different steady-state creep mechanisms under creep loading process. For example, GB diffusion is the dominant creep mechanism of metals as the stress and grain size exponents $n$ and $p$ is 1 and 3, respectively[43]. Here, the insight is further developed to describe the creep behaviors of polycrystalline methane hydrates. In previous studies [19, 20], stress exponents $n$ is also used to indicate the creep flow mechanisms of polycrystalline methane hydrates in terms of the standard high-temperature creep constitutive relationship[21]. As shown in Figure 6a, grain size exponent $p$ firstly nonlinearly decreases from around 3.46 to 0.32, and then it changes from 0.32 to -0.92. This change from positive to negative represents the transition of mechanical creep mechanisms. For example, under loading stress of 300 MPa at 263.15 K, creep mechanism of polycrystalline methane hydrates with small grain size is mainly dominated by crystal lattice vibration and GB diffusion. Whereas for polycrystals with large grain size, the creep mechanism is primarily governed by GB diffusion/sliding. It is observed from Figure 6b that stress exponent $n$ of polycrystalline methane hydrates with various grain sizes nonlinearly increases from a low to a high value with increasing applied static stress. The creep mechanism of polycrystalline methane hydrates can be also indicated from the values of stress exponent $n$. As the applied static stress increases, the main creep mechanisms of polycrystals with large grain size change from crystal lattice vibration and GB diffusion to GB diffusion/sliding. With regard to polycrystalline samples with small grain size, the creep mechanisms are mainly crystal lattice vibration and GB diffusion at low/high temperature, instead of GB sliding. Therefore, for a given polycrystalline methane hydrate sample, the creep mechanism is mainly controlled by several factors, namely, grain size, static loading stress, and temperature. The combinations of grain size exponent $p$, stress exponent $n$ and temperature can be utilized to indicate the steady-state creep mechanisms. Those are analogous to the experimental and numerical simulation results of other solid materials[47, 48]. Both ice and gas hydrates are mainly composed of water molecules connected by hydrogen-bonds. Similarly to conventional metallic substances at high temperature, the creep manner of both polycrystalline ice and hydrates is dependent on the applied static stress and temperature[49, 50]. Both polycrystals show a much more remarkable creep behavior with increasing the static stress and the increasing temperature. In constant-load creep tests, gas hydrate-consolidated sands are much stronger than ice-consolidated sands, exhibiting noticeable differences in their creep behaviors[51].

**Conclusions**

In summary, mechanical creep behaviors of polycrystalline methane hydrates are systematically investigated using large-scale MD simulations with coarse-grained water model. It is found that mechanical creep responses of nanocrystalline methane hydrates are strongly affected by crystalline grain size, external conditions of temperature and static stress. There are three creep deformation stages identified from the creep responses. Particularly, the secondary creep deformation stage is characteristic long steady-state creep. The mechanical creep responses of nanocrystalline methane hydrates are primarily dominated by deformation of crystalline grains, GB diffusion and GB sliding. Remarkably, structural transformation also plays an important role in the mechanical creep mechanisms of nanocrystalline methane hydrates. The findings are important to understand the mechanical stability and evolution scenarios of naturally occurring gas hydrate reservoirs.

## ASSOCIATED CONTENT

**Supporting Information**

The Supporting Information is available free of charge at https://xxx.

## AUTHOR INFORMATION

**Corresponding Author**

*pinqiang@wust.edu.cn;

**Notes**

The authors declare no competing financial intertest.

**Acknowledgments**

This work was financially supported by Starting Research Fund from Wuhan University of Science and Technology (Grant No. 1160034), the National Natural Science Foundation of China (Grant No. 11772278), the Jiangxi Provincial Outstanding Young Talents Program (Grant No. 20192BCBL23029), the Fundamental Research Funds for the Central Universities (Xiamen University: Grant Nos. 20720180014 and 20720180018).

**References**

1. B. B. Rath, *MRS Bull.*, 2011, **33**, 323-325.

2. M. R. Walsh, S. H. Hancock, S. J. Wilson, S. L. Patil, G. J. Moridis, R. Boswell, T. S. Collett, C. A. Koh and E. D. Sloan, *Energy Economics*, 2009, **31**, 815-823.

3. K. A. Kvenvolden, *Reviews of Geophysics*, 1993, **31**, 173-187.

4. E. D. Sloan, *Nature*, 2003, **426**, 353-363.

5. C. A. Koh, *Chem. Soc. Rev.*, 2002, **31**, 157-167.

6. E. G. Hammerschmidt, *Ind. Eng. Chem.*, 1934, **26**, 851-855.

7. S. P. Hesselbo, *Nature*, 2000, **406**, 392-395.

8. K. Lambeck, T. M. Esat and E. K. Potter, *Nature*, 2002, **419**, 199-206.

9.  B. J. Phrampus and M. J. Hornbach, *Nature*, 2012, **490**, 527-530.

10. H. Lee, J.-w. Lee, D. Y. Kim, J. Park, Y.-T. Seo, H. Zeng, I. L. Moudrakovski, C. I. Ratcliffe and J. A. Ripmeester, *Nature*, 2005, **434**, 743-746.

11. R. E. Rogers and Y. U. Zhong, *Ann. N.Y. Acad. Sci.*, 2000, **912**, 843-850.

12. J. S. Gudmundsson, V. Andersson, O. I. Levik and M. Mork, *Ann. N.Y. Acad. Sci.*, 2000, **912**, 403-410.

13. S. L. Miller and W. D. Smythe, *Science*, 1970, **170**, 531.

14. H. Lu, Y.-t. Seo, J.-w. Lee, I. Moudrakovski, J. A. Ripmeester, N. R. Chapman, R. B. Coffin, G. Gardner and J. Pohlman, *Nature*, 2007, **445**, 303.

15. K. A. Kvenvolden, *Chem. Geol.*, 1988, **71**, 41-51.

16. J. S. Loveday, *Nature*, 2001, **410**, 661-663.

17. J. I. Lunine and D. J. Stevenson, *Astrophys. J. Suppl.*, 1985, **58**, 493-531.

18. M. D. Max and S. M. Clifford, *J. Geophys. Res.-Planets.*, 2000, **105**, 4165-4171.

19. W. B. Durham, L. A. Stern and S. H. Kirby, *Can. J. Phys.*, 2003, **81**, 373-380.

20. W. B. Durham, S. H. Kirby, L. A. Stern and W. Zhang, *J. Geophys. Res.*, 2003, **108**, 2182.

21. J.-P. Poirier, *Creep of Crystals: High-Temperature Deformation Processes in Metals, Ceramics and Minerals*, Cambridge University Press, Cambridge, 1985.

22. K. Miyazaki, N. Tenma and T. Yamaguchi, *Energies*, 2017, **10**, 1466.

23. G. F. Voronoi, *Journal für die reine und angewandte Mathematik*, 1908, **134**, 198-287.

24. L. A. Stern, S. H. Kirby and W. B. Durham, *Energy Fuels*, 1998, **12**, 201-211.

25. F. H. Stillinger and T. A. Weber, *Phys. Rev. B*, 1985, **31**, 5262-5271.

26. V. Molinero and E. B. Moore, *J. Phys. Chem. B*, 2009, **113**, 4008-4016.

27. L. C. Jacobson and V. Molinero, *J. Phys. Chem. B*, 2010, **114**, 7302-7311.

28. J. Wu, F. Ning, T. T. Trinh, S. Kjelstrup, T. J. H. Vlugt, J. He, B. H. Skallerud and Z. Zhang, *Nat. Commun.*, 2015, **6**, 8743.

29. Y. Bi and T. Li, *J. Phys. Chem. B*, 2014, **118**, 13324-13332.

30. Y. Bi, A. Porras and T. Li, *J. Chem. Phys.*, 2016, **145**, 211909.

31. S. J. Plimpton, *J. Comput. Phys.*, 1995, **117**, 1-19.

32. P. Cao, F. Ning, J. Wu, B. Cao, T. Li, H. A. Sveinsson, Z. Liu, T. J. H. Vlugt and M. Hyodo, *ACS Appl. Mater. Interfaces*, 2020, **12**, 14016-14028.

33. L. A. Stern, W. B. Durham and S. H. Kirby, *J. Geophys. Res.-Solid Earth*, 1997, **102**, 5313-5325.

34. P. Cao, J. Wu, Z. Zhang and F. Ning, *Nanotechnology*, 2017, **28**, 045702.

35. J. Schiøtz, F. D. D. Tolla and K. W. Jacobsen, *Nature*, 1998, **391**, 561-563.

36. J. Wu, P. Cao, Z. Zhang, F. Ning, S.-s. Zheng, J. He and Z. Zhang, *Nano Lett.*, 2018, **18**, 1543-1552.

37. M. F. Ashby and D. R. H. Jones, *Engineering materials 1: an introduction to properties, applications and design*, Elsevier, London, 2012.

38. A. K. Mukherjee, J. E. Bird and J. E. Dorn, *Experimental correlations for high-temperature creep*, California Univ., Berkeley. Lawrence Radiation Lab., United States, 1968.

39. J. Vatamanu and P. G. Kusalik, *J. Am. Chem. Soc.*, 2006, **128**, 15588-15589.

40. M. R. Walsh, C. A. Koh, E. D. Sloan, A. K. Sum and D. T. Wu, *Science*, 2009, **326**, 1095-1098.

41. S. Liang and P. G. Kusalik, *J. Chem. Phys.*, 2015, **143**, 011102.

42. Z. Zhang, G.-J. Guo, N. Wu and P. G. Kusalik, *J. Phys. Chem. C*, 2020, DOI: 10.1021/acs.jpcc.0c07375.

43. R. L. Coble, *J. Appl. Phys.*, 1963, **34**, 1679-1682.

44. J. Bardeen, *Phys. Rev.*, 1949, **76**, 1403-1405.

45. H. Lüthy, R. A. White and O. D. Sherby, *Mater. Sci. Eng.*, 1979, **39**, 211-216.

46. W. Blum and X. H. Zeng, *Acta Mater.*, 2009, **57**, 1966-1974.

47. W. M. Yin, S. H. Whang, R. Mirshams and C. H. Xiao, *Mater. Sci. Eng., A*, 2001, **301**, 18-22.

48. N. Wang, Z. Wang, K. T. Aust and U. Erb, *Mater. Sci. Eng., A*, 1997, **237**, 150-158.

49. J. W. Glen, *Proc. R. Soc. Lond. A*, 1955, **228**, 519-538.

50. J. Weertman, *Ann. Rev. Earth Planet. Sci.*, 1983, **11**, 215-240.

51. I. Cameron, Y. P. Handa and T. H. W. Baker, *Can. Geotech. J.*, 1990, **27**, 255-258.

**Figures and Captions**

**Figure 1**

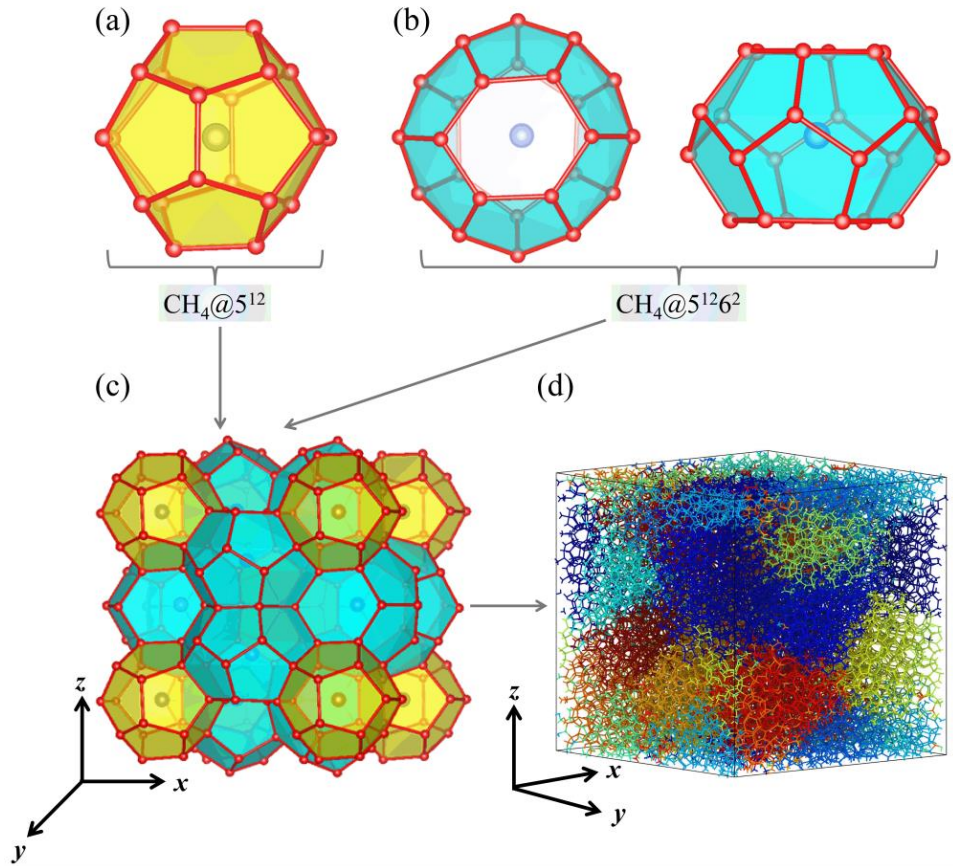

**Figure 1. Molecular structures of methane hydrate.** (a) Small polyhedral water cage ($5^{12}$) and (b) large polyhedral water cage ($5^{12}6^2$). (c) A cluster of gas hydrate made of both large ($5^{12}6^2$) and small cage ($5^{12}$). (d) Three-dimensional (3D) polycrystalline methane hydrates with an average grain size of 4.0 nm. Side view of a typical polycrystal in which crystalline grains are colored for clarification of grain morphology.

**Figure 2**

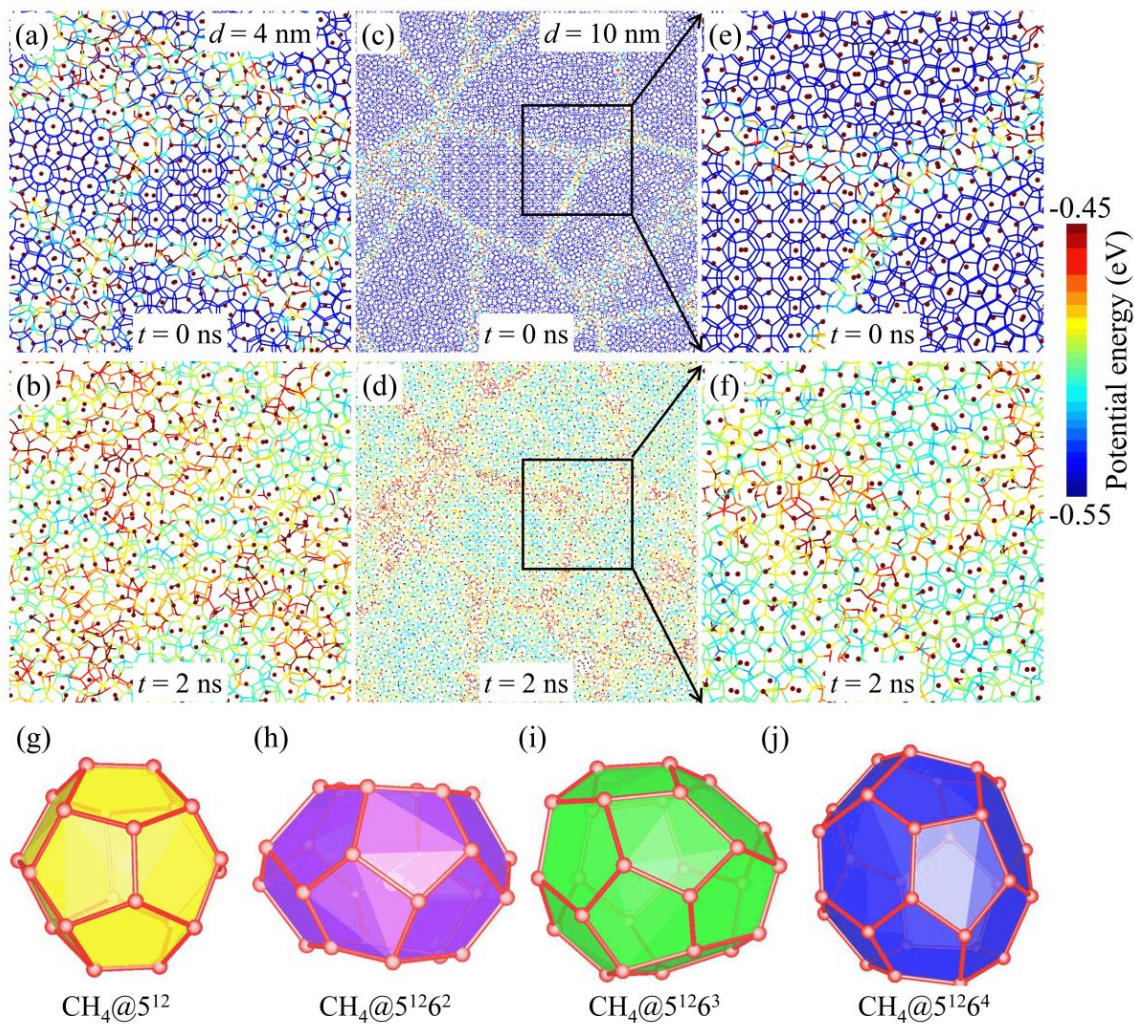

**Figure 2. Cross-sectional structures of methane hydrate polycrystals.** (a)-(b) Cross-sectional snapshots of polycrystals with an average grain size of 4.0 nm at their initial and relaxed states, respectively. (c)-(d) Cross-sectional snapshots of polycrystals with an average grain size of 10.0 nm before and after relaxed states, respectively. (e)-(f) Zoomed-in configurations of grain boundaries marked by black rectangles in (c) and (d) are highlighted, respectively. All molecules are rendered according to their potential energy. (g)-(j) The water-based clathrate cages of (g)

$CH_4@5^{12}$, (h) $CH_4@5^{12}6^2$, (i) $CH_4@5^{12}6^3$ and (j) $CH_4@5^{12}6^4$ are captured at the grain boundaries, respectively.

**Figure 3**

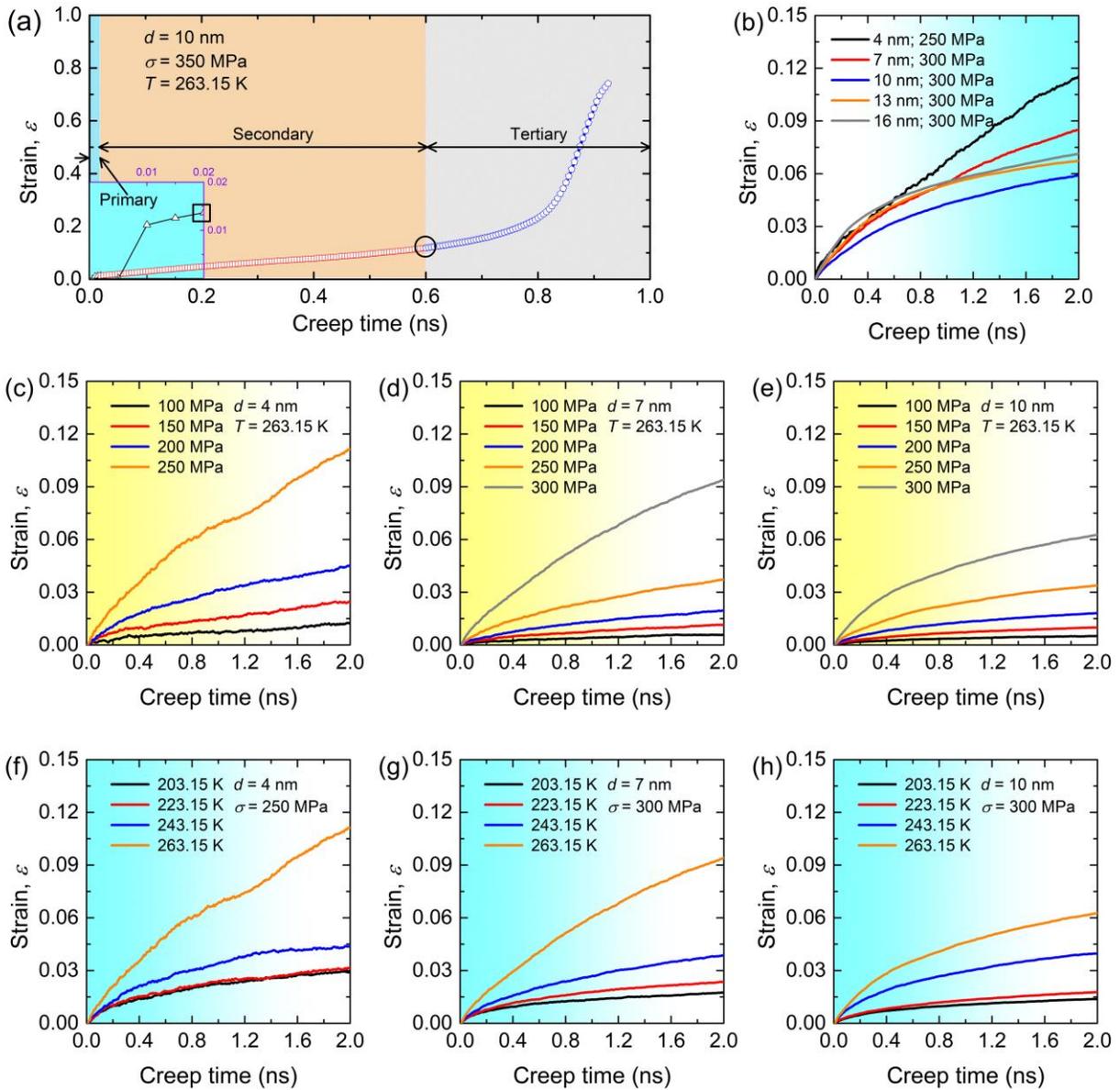

**Figure 3. Creep behaviors of polycrystalline methane hydrates.** (a) Creep strain - simulation time curves for polycrystalline samples with an average grain size of 10 nm under loading stress

of 350 MPa at temperature of 263.15 K. Three different creep stages from the creep curves are identified as indicated by the colored backgrounds. The inset figure is a zoomed-in view of Figure (a) to clearly show the first primary creep response stage. (b) Creep strain - MD simulation time curves of polycrystalline methane hydrates with different average grain sizes. (c)-(e) Creep strain - MD simulation time curves of polycrystals with an average grain size of 4.0, 7.0 and 10.0 nm at temperature of 263.15 K subjected to different loading stresses, respectively. (f)-(h) Creep strain - MD simulation time curves of polycrystals with an average grain size of 4.0, 7.0 and 10.0 nm under a given load at different temperatures, respectively.

**Figure 4**

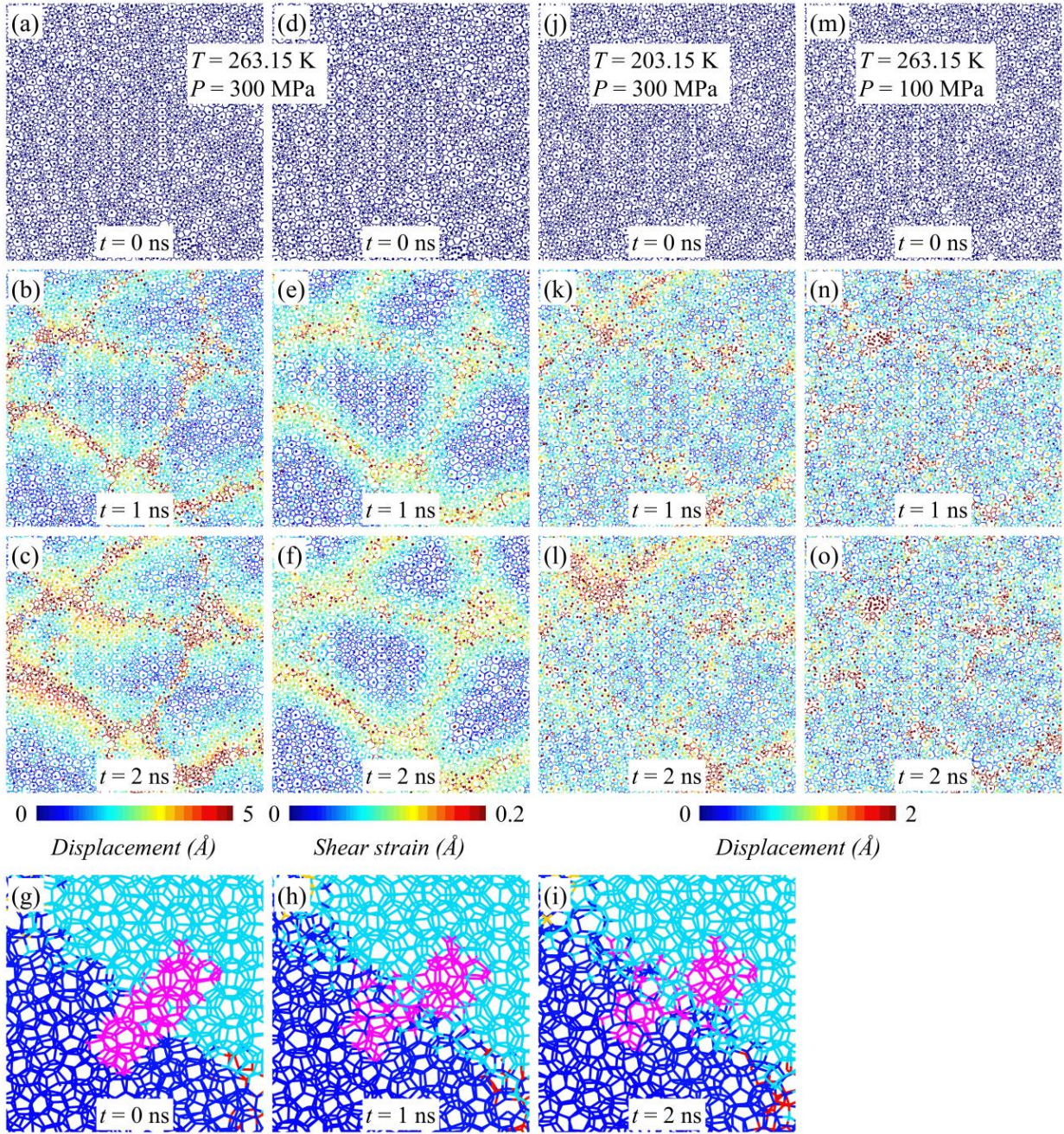

**Figure 4. Localized structures of polycrystalline methane hydrates.** (a)-(c) Cross-sectional snapshots of polycrystals with an average grain size of 10 nm under load of 300 MPa at temperature of 263.15 K at creep times of 0 ns, 1 ns, and 2 ns, respectively. The snapshots are colored according to the molecular displacements relative to the initial state. (d)-(f) Cross-sectional snapshots of polycrystals with an average grain size of 10 nm under loading stress of 300 MPa at temperature of 263.15 K at creep times of 0 ns, 1 ns, and 2 ns, respectively. The snapshots are colored on the basis of the values of shear strain. (g)-(i) Cross-sectional snapshots of polycrystals with an average grain size of 10 nm under stress level of 300 MPa and temperature of 263.15 K at creep times of 0 ns, 1 ns, and 2 ns, respectively. Grains are differently colored. Molecules falling in a unique region is magenta-colored for monitoring grain boundary sliding. (j)-(l) Cross-sectional snapshots of polycrystals with an average grain size of 10 nm under loading stress of 300 MPa at temperature of 203.15 K at creep times of 0 ns, 1 ns, and 2 ns, respectively. The snapshots are colored according to the molecular displacements relative to the initial state. (m)-(o) Cross-sectional snapshots of polycrystals with an average grain size of 10 nm under loading stress of 100 MPa at temperature of 263.15 K at creep times of 0 ns, 1 ns, and 2 ns, respectively. The snapshots are colored according to the molecular displacements relative to the initial state.

**Figure 5**

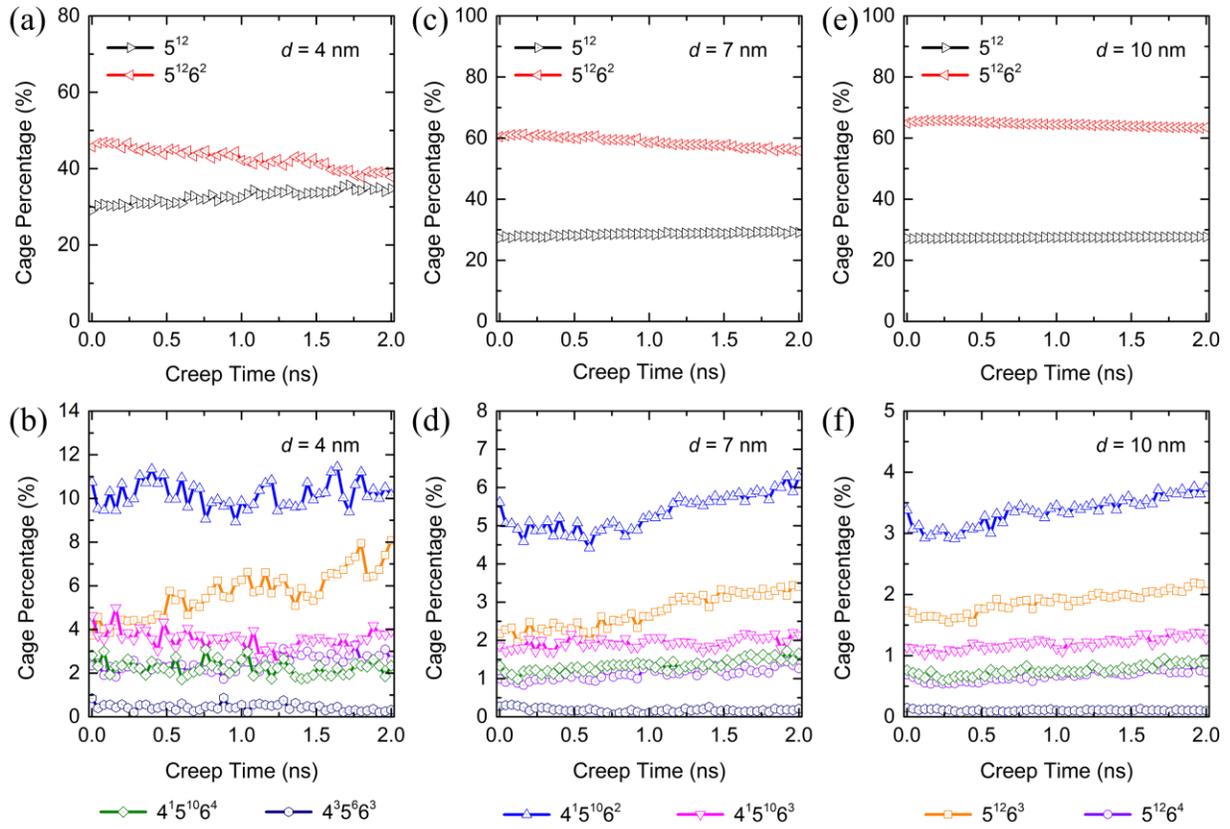

**Figure 5. The percentages of identified water cages in nanocrystalline methane hydrates during mechanical creep loads.** (a)-(b) nanocrystalline methane hydrates with an average grain size of 4 nm under loading stress of 250 MPa at 263.15 K. (c)-(d) nanocrystalline methane hydrates with an average grain size of 7 nm under loading stress of 300 MPa at 263.15 K. (e)-(f) nanocrystalline methane hydrates with an average grain size of 10 nm under loading stress of 300 MPa at 263.15 K. The percentage of each cage type is obtained via dividing the number of each identified cage type by the total number of all identified cage types in this work.

**Figure 6**

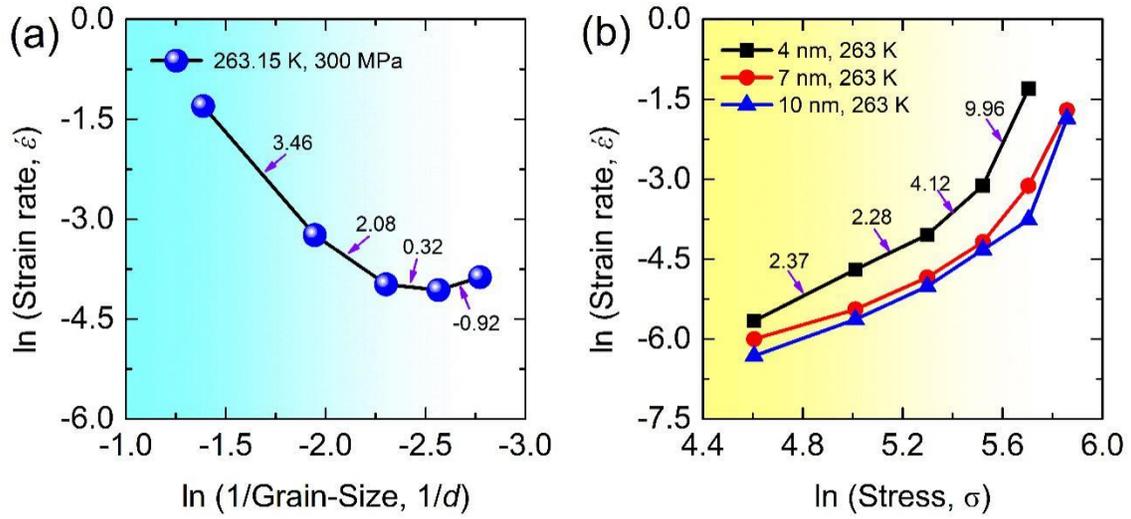

**Figure 6.** (a) Logarithmic plot of steady-state creep strain rate versus inverse of grain size for polycrystalline methane hydrates under loading stress of 300 MPa at 263.15 K. (b) Logarithmic plots of steady-state creep strain rate versus stress at 263.15 K for polycrystalline methane hydrates with mean grain size of 4.0, 7.0 and 10.0 nm. The numbers marked by the purple arrows represent the grain size exponent $p$ and stress exponent $n$, respectively.